\documentclass[aps,prl,twocolumn,amsmath,amssymb,superscriptaddress, reprint]{revtex4-1}
\usepackage{enumerate}
\usepackage{amsmath}
\usepackage{graphicx}
\usepackage{color}
\usepackage[breaklinks]{hyperref}
\usepackage{psfrag}
\usepackage{stmaryrd}
\usepackage{amssymb}
\usepackage[latin1]{inputenc}
\usepackage[english]{babel}
\usepackage{float}
\def \degree {^\mathrm{o}}
\let\oldtabular=\tabular
\def\tabular{\small\oldtabular}
\usepackage{mathrsfs}
\usepackage[margin=3cm]{geometry}
\usepackage{array}

\def \degree {^\mathrm{o}}

\begin{document}

\author{V. De Zotti}
\affiliation{Universit\'e de Lyon, ENSL, UCBL, CNRS, Laboratoire de Physique, Lyon, France}

\author{K. Rapina}
\affiliation{Universit\'e de Lyon, ENSL, UCBL, CNRS, Laboratoire de Physique, Lyon, France}

\author{P.-P. Cortet}
\affiliation{Laboratoire FAST, CNRS, Universit\'e  Paris-Sud, Universit\'e Paris-Saclay, Orsay, France}

\author{L. Vanel}
\affiliation{Université de Lyon, Université Claude Bernard Lyon 1, CNRS, Institut Lumière Matière, F-69622, Villeurbanne, France}

\author{S. Santucci}
\email{stephane.santucci@ens-lyon.fr}
\affiliation{Universit\'e de Lyon, ENSL, UCBL, CNRS, Laboratoire de Physique, Lyon, France}
\affiliation{Lavrentyev Institute of Hydrodynamics, Novosibirsk, Russia}

%\date{\today}
%\title{Transfer of bending to kinetic energy controls adhesive peel front micro-instability}
\title{Bending to kinetic energy transfer in adhesive peel front micro-instability}

\begin{abstract}
We report an extensive experimental study of 
a detachment front dynamics instability, appearing at microscopic scales during the peeling of adhesive tapes.
The
amplitude
of this instability
scales with its period as $A_{\text{mss}} \propto T_{\text{mss}}^{1/3}$,  
with a pre-factor
evolving slightly with the peel angle $\theta$, and increasing systematically  with the bending modulus $B$ of the tape backing. 
Establishing a local energy budget of the detachment process during one period of this micro-instability, 
our theoretical model shows that the
elastic bending
energy stored in the portion of tape to be peeled  is 
converted into 
kinetic energy,  
providing a quantitative description of
the experimental scaling law. 
\end{abstract}

%\pacs{62.20.mm,68.35.Np,82.35.Gh}
%62.20.mm Fracture
%68.35.Np Adhesion
%82.35.Gh Polymers on surfaces; adhesion

\maketitle

The periodic velocity oscillations of the detachment front during the peeling of adhesive tapes constitute 
an archetypal example of a dynamical rupture instability. 
This  stick-slip motion leads to a screechy sound that everyone has experienced, when peeling-off packing tape.  However, despite a large number of studies \cite{Gardon1963,Aubrey1969,Racich1975,Barquins1986,Maugis1988,Barquins1997,Ryschenkow1996,Gandur1997,Ciccotti2004,De2004,Cortet2007,De2008}, this instability is not fully understood and still causes industrial problems, with deafening noise levels, and damages to both 
adhesives and peel systems.

The effective fracture energy of adhesive-substrate joints can decrease 
over certain ranges of peel front velocity~\cite{,Aubrey1969,Racich1975,Barquins1986,Maugis1988}. 
In such unstable condition, for which less energy is required for the crack front to grow faster, a transition from a quasi-static rupture mode to a dynamic one occurs, as for frictional interfaces~\cite{Baumberger1994, Rubinstein2004, Svetlizky2014}. 
During the rapid slip phases,  
the dynamical mode of failure is likely to give rise to small scale spatio-temporal front instabilities~\cite{Fineberg1999}.
Indeed, ultra-fast imaging could unveil that the peel front locally advances by steps in the main peel direction as a result of the propagation of a dynamic fracture kink in the transverse direction, at spatio-temporal scales much smaller than the macroscopic stick-slip~\cite{Thoroddsen2010, Marston2014, Dalbe2015}: 
 the kink occurs periodically at ultrasonic frequencies with an amplitude of a few hundred microns, %This is observed 
 not only during the slip phase of the macro-instability, but also, for imposed peel velocities in a finite range beyond the macro-stick-slip domain where the peeling is regular at macroscopic scales~\cite{Dalbe2015}.

Interestingly, this micro-instability of the peel front characterized by the side-ways propagation of fracture kinks share similarities with other physical processes, as for instance the local contact lines dynamics on textured surfaces \cite{Gauthier2013}, or the dislocations motion in the 
yielding of crystalline materials \cite{Dislocations}.  
While it was shown that those transverse cracks are accompanied by cycles of load and release of the elastic bending energy stored in the tape backing  in the vicinity of the peel front \cite{Dalbe2015}, the physical origin of the micro-instability and its interaction with the macroscopic one remains an open issue.

In this Letter, 
we provide a detailed experimental study of 
this micro-instability, varying systematically the peeled length $L$, 
the peel angle $\theta$, 
the lineic mass $\mu$ and bending modulus $B$ of the ribbon, over a wide range of driving peel
velocities $V$. 
Thanks to a large data statistics, we show that
the  micro-instability amplitude scales with its period as $A_{\text{mss}} \propto T_{\text{mss}}^{1/3}$,  
with a pre-factor that increases 
with the bending
modulus of the tape backing. We demonstrate that  
 the bending elastic energy of the ribbon released during each micro-slip is converted into kinetic energy, 
 allowing a quantitative prediction of this scaling law.

We peel a 3M Scotch$^{\mbox{\scriptsize{\textregistered}}}$ 600
adhesive tape from a transparent flat substrate by winding its
extremity at a constant velocity $V$ with a brushless motor.
Changing the relative position of the substrate and peeling motor, we can easily vary the peel angle $\theta$ and ribbon length
$L$. The peeling  
of this 
tape (a
polyolefin blend backing $b=19~$mm wide, $e=34~\mu$m thick,
tensile modulus $E=1.41~$GPa, density $\mu=8.10^{-4}~$kg/m) coated
with a 15~$\mu$m  synthetic acrylic adhesive layer has 
been widely studied~\cite{Barquins1986, Maugis1988,Barquins1997,
Thoroddsen2010, Cortet2013, Marston2014, Dalbe2014, Dalbe2015,
Dalbe2014b, Dalbe2016}.
In contrast with those  
studies, for each experiment,
using a scalpel, we carefully extract  
two layers of the
adhesive tape from its original roller. We attach them 
on a transparent plate and then, perform the peeling at the interface
between those two layers. 
The release side of
 the adhesive tape backing constitutes the top part of the
substrate of our peel experiments. 
Thanks to the homogeneous properties of the commercial roller adhesive layer,  this protocol improves the reproducibility of our experiments, by avoiding a pre-peeling which may damage the adhesive. 
\begin{figure}[h!]
\centerline{\includegraphics[width=7.5cm]{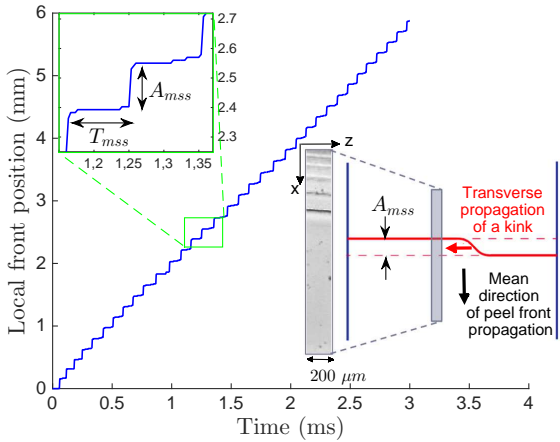}}
\vspace{-0.3cm} 
\caption{ Micro-stick-slip dynamics of the local longitudinal position of the peel front during an experiment at $V=1.8$~m/s,
$L=50$~cm and $\theta =90 \degree$, with periods of rests $T_{\text{mss}}$ 
preceding slips of size $A_{\text{mss}}$, as a result of the transverse propagation across the tape of a kinked fracture of amplitude $A_{\text{mss}}$. We also display a typical image recorded by the fast camera (the grey zone in inset gives its reduced field of view).}
\label{fig:Example}
\end{figure}

A Photron SA5  fast camera with a macro-lens images a small portion of 
the peel front
through the transparent substrate with a 
resolution of
9.8~$\mu$m/pixel, at a 
rate of $175\,000$ or
$300\,000$~fps, % 
for $640 \times 56$~px$^2$ and $832 \times
24$~px$^2$ images respectively. 
Analyzing the grey levels of each image, we extract the detachment
front longitudinal position $x$ at a given transverse position $z$. 
During an acquisition, this
front typically advances of a few mm, so that $L$ and $\theta$
 can be
considered constant (varying less than $5\%$). 
A sketch of the setup can be found in supplemental material together with typical recorded videos.

In the driving velocity range $[V_a, V_d]$~\cite{thresholdmacro}, for which the peel force decreases, the detachment front displays the classical stick-slip instability, 
 with regular velocity oscillations at the millisecond timescale,
and millimetric slips related to cycles of loading and release of the stretching energy stored in the whole peeled tape.
The oscillating front velocity is thus different from the driving velocity $V$ imposed at the extremity of the peeled tape, which is the control parameter of the experiment. 
When the mean front velocity (measured at the ms timescale) becomes larger than 
$v_a \simeq 1$~m/s~\cite{DeZotti2018}, the peel front advances by the  propagation  of fracture kinks, across the tape width, in the transverse direction $z$, at very high velocities, 
 from $650$~m~s$^{-1}$ up to $900$~m~s$^{-1}$~\cite{Dalbe2015}. 
 Those dynamic transverse cracks occur during the slip phase of the macro-instability, but also, permanently as shown in Fig.~\ref{fig:Example}, for driving velocities  above $V_d$, the disappearance threshold of the macro-instability.
 This micro-instability finally disappears as well, when the mean peel front velocity is above $v_d \simeq 20$~m/s \cite{Dalbe2015}.
Figure~\ref{fig:Example} gives a typical example of the local
front position time series for an experiment at $V=1.8 $~m/s, $L=50$~cm, $\theta =90 \degree$. 
While the average front velocity measured at the ms timescale 
is equal to the driving velocity $V$ %(no macro-stick-slip),  
(regular peeling without macro-stick-slip),  
at shorter timescales, we observe a staircase dynamics
with sudden jumps of amplitude $A_{\text{mss}}\simeq 170$~$\mu$m separated
by periods of rest of $T_{\text{mss}}\simeq 100$~$\mu$s.  

\begin{figure}[h!]
\centerline{\includegraphics[width=8.4cm]{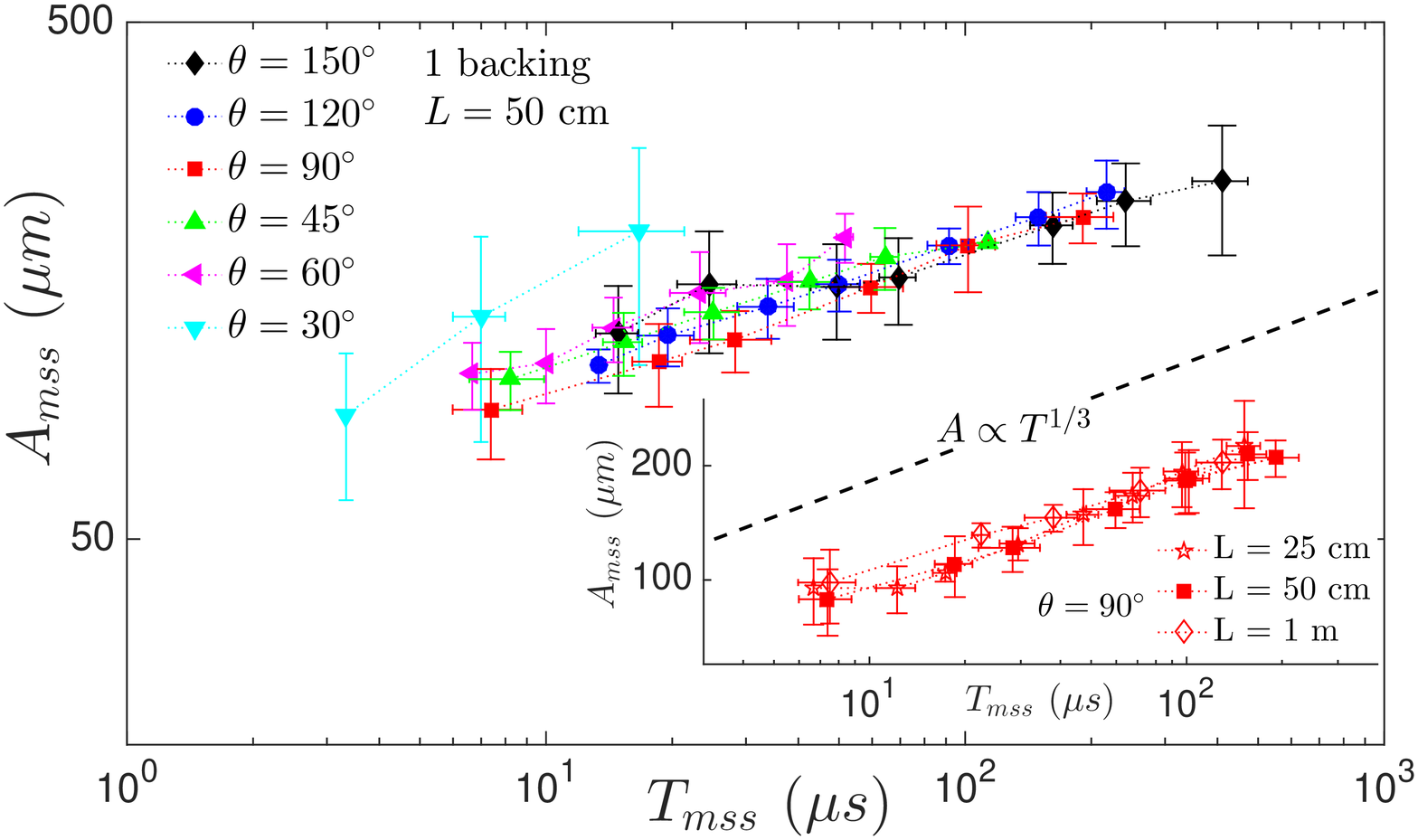}}
\vspace{-0.05cm}
\centerline{\includegraphics[width=8.4cm]{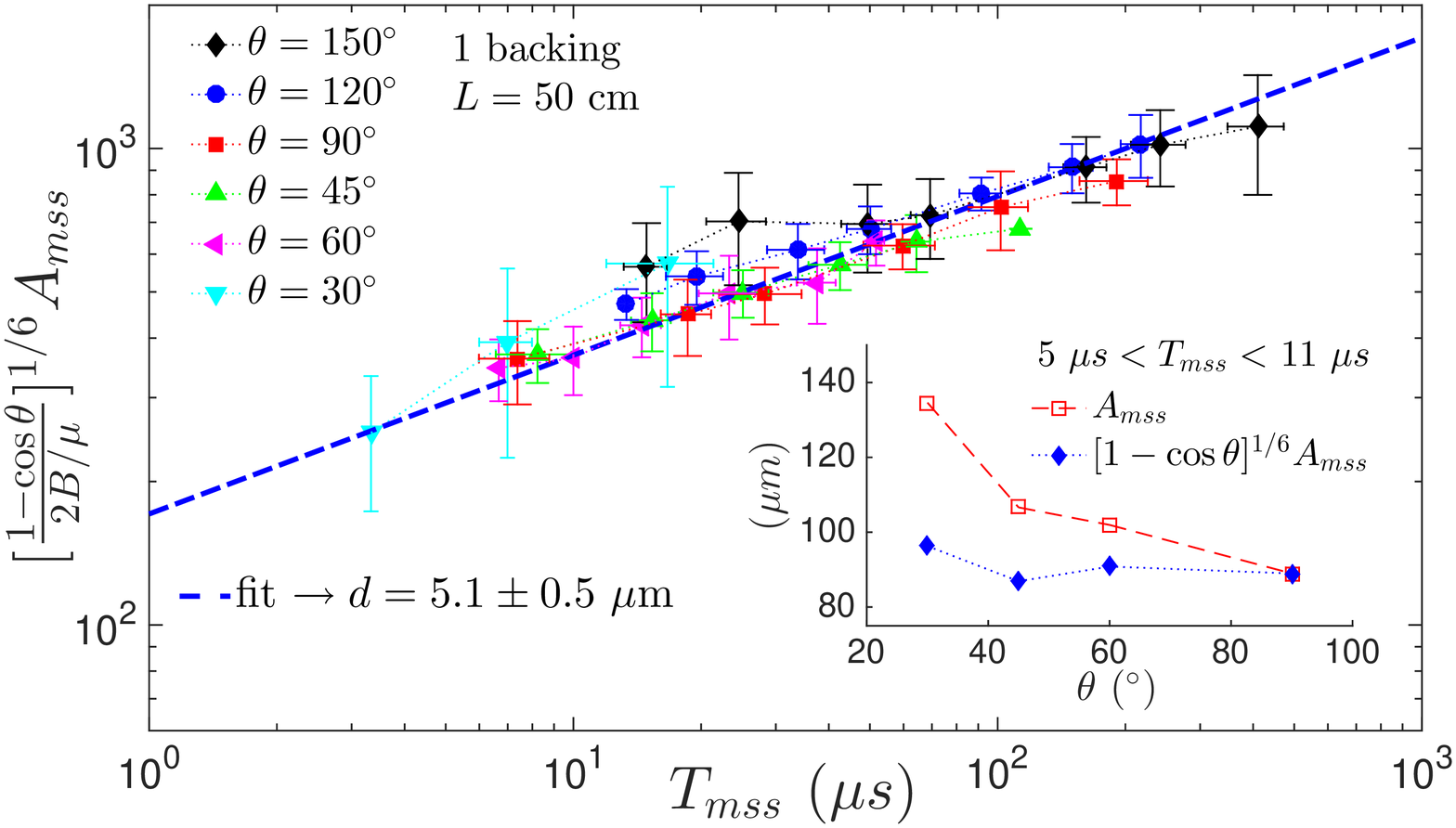}}
\vspace{-0.35cm} \caption{Mean amplitude of the micro
stick-slips $A_{\text{mss}}$ as a function of the mean duration
$T_{\text{mss}}$ for a wide range of experimental conditions at
different peel velocities: different peel angle $\theta$ at
$L=50$~cm and for $\theta = 90\degree$ and three peeled lengths $L$ (inset). Bottom panel shows the same data but
with $A_{\text{mss}}$ rescaled by $\left[ 2B/ (\mu
(1-\cos\theta))\right]^{1/6}$ following Eq.~(\ref{eq:AT}). The inset shows that the micro-slips amplitude of periods 
$ T_{\text{mss}} \in [5,11]~\mu$s  evolve as  $(1-\cos \theta)^{-1/6}$.}\label{fig:L_Theta_B}
\end{figure}

From  the peel front temporal evolution measured for
each experiment, we could detect several thousands of
micro-stick-slip events. 
For each of them, we extract the period of rest  
$T_{\text{mss}}$ 
preceding a micro-slip of amplitude  $A_{\text{mss}}$. 
In Fig.~\ref{fig:L_Theta_B} (top), we display   
$A_{\text{mss}}$  as a function of 
$T_{\text{mss}}$,  averaged in logarithmic bins. 
We gather here data of numerous experiments:
for a fixed peel angle $\theta = 90\degree$ and several peeled
ribbon length $L$ (inset), and changing 
this angle $\theta$ while keeping $L$ fixed to $50$~cm (main panel). 
We clearly observe that
$A_{\text{mss}}$ increases with $T_{\text{mss}}$, following a
power-law scaling with an exponent close to $1/3$, 
independent of the peeled length $L$ (inset).  On the
other hand, both the range and pre-factor of the scaling law
evolve slightly with the peel angle  (inset, bottom panel), but seemingly not the
power-law exponent. 

\begin{figure}[h!]
\centerline{\includegraphics[width=8.4cm]{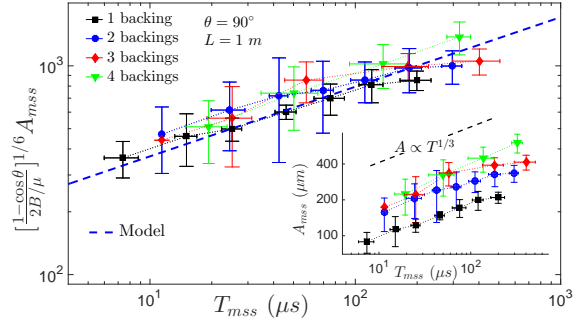}}
\vspace{-0.4cm} \caption{
Mean amplitude of the micro
stick-slips $A_{\text{mss}}$ as a function of the mean duration
$T_{\text{mss}}$ for peel experiments at
$L=1$~m and $\theta = 90\degree$, with tape of different ribbon thicknesses measured in
number of superimposed glued backings. 
}\label{fig:BK}
\end{figure}

In order to evaluate the role of the  elastic bending energy,
stored locally in the vicinity of the peel front, 
we have 
studied the impact of the ribbon bending modulus $B$. 
Superimposing up to 4
layers of the 3M Scotch$^{\mbox{\scriptsize{\textregistered}}}$
600 rigid backing (cleansed of its adhesive coating) with a rigid
glue (using $ e_L = 26\pm10$ $\mu$m layers of
Loctite$^{\mbox{\scriptsize{\textregistered}}}$ 406), we could
increase the bending modulus of our adhesive tape by 2 orders
of magnitude. 
Indeed, considering that the n-layer backing of thickness $h_n = n e + (n-1) e_L$  has
the same tensile modulus $E$ than the 3M
Scotch$^{\mbox{\scriptsize{\textregistered}}}$ 600 ribbon, 
its bending modulus can be estimated as $B = Eh_n^3 b /12$.
As a result, we find a systematic increase in the micro-slips size 
$A_{\text{mss}}$ with  
$B$, as  
shown in
Fig.~\ref{fig:BK} (inset), preserving nevertheless the
$T_{\text{mss}}^{1/3}$ scaling. 

We discuss now the evolution of the adhesive tape close to the peel front, in order to obtain a local energy balance of the detachment process during a microscopic stick-slip cycle (per unit width of the peel front). 
During the micro-stick phase, the adhesive layer is stretched, while the ribbon is bent locally at the vicinity of the static detachment front. 
We assume that 
a micro-slip of size $A_{\text{mss}}$ occurs suddenly, when the %adhesive layer 
glue is stretched up to a critical length $d$. During this fast interfacial crack propagation, both the elastic bending energy of the ribbon $E_{B-r}$ and the %elastic 
stretching energy of the glue $E_{S-g}$ stored during the micro-stick phase are released.  
Therefore, we can write the  local energy balance during a micro-stick-slip: %cycle:
\begin{equation}\label{eq:EB}
       E_{B-r} + E_{S-g} = \Gamma A_{\text{mss}} + E_k, 
\end{equation}
where $\Gamma$ is the effective fracture energy of the adhesive-substrate joint, and, 
$E_k$ corresponds to the excess of elastic energy released, converted into kinetic energy that the tape locally acquires. 
Such energy balance is similar to the one proposed by Mott \cite{Mott1948}, generalizing Griffith's criterion \cite{Griffith1920} to describe dynamic rupture processes.
%The effective fracture energy  $\Gamma$ takes into account all dissipative processes occurring when the glue detaches from the substrate. It is much larger than the interfacial energetic cost  $\Gamma_i$ of creating new surfaces, so that,  $\Gamma =  \Gamma_h +  \Gamma_i  \simeq  \Gamma_h$. 
%
%

The effective fracture energy $\Gamma$   has been theoretically related to the energy needed to deform the adhesive layer up to a critical strain at which it debonds~\cite{Kaelbe1964,Gent1969,Villey2015}. 
Recent experiments with polyacrylate adhesives have shown that this energy is indeed proportional to the integral of the non-linear rate-dependent stress-strain curve of the confined glue~\cite{Chopin2018}.
The critical elongation of the glue at debonding $d$ is thus a crucial parameter in the determination of the peeling energy 
 $\Gamma$  (the value of $d$ and its dependencies with material, geometrical and dynamical parameters is still an open issue~\cite{Chopin2018,Villey2017,Creton2016}).
 In this context, we can consider that the energy accumulated in the deformation of the glue  $E_{S-g}$ is completely dissipated  when it debonds and a micro-slip of size $A_{\text{mss}}$ occurs, i.e. $\Gamma A_{\text{mss}}=E_{S-g}$.
As a consequence, the local energy balance (\ref{eq:EB}) leads to the transfer of the ribbon bending energy 
$E_{B-r}$ 
into an increase of kinetic energy $E_k$ that the tape locally gains during a micro-slip, 
       $E_{B-r} =  E_k$.

Assuming that the tape is bent over a length scale equal to the micro-slip size  $A_{\text{mss}}$, just before its triggering at the critical elongation of the glue $d$, with a radius of curvature $R_0 \simeq A_{\text{mss}}^2/2d$, 
 the order of magnitude of the bending energy released during a micro-slip writes $E_{B-r} \simeq \frac12 B A_{\text{mss}} / R_0^2  \simeq 2 B d^2/A_{\text{mss}}^3 $. 
In our model, the micro-slip duration is fixed by the time scale for the release of this local bending energy $E_{B-r}$,  
 controlled, a priori, by the bending waves period 
 $\tau=\frac{\lambda^2}{2\pi}  \sqrt{\frac{\mu}{B}}$~\cite{Landau}  
 of wavelength $\lambda = A_{\rm mss}$. 
 Considering a single tape layer with  $\lambda\sim 150 - 200~\mu$m gives a time scale $\tau\sim 0.3-0.6~\mu$s, 
 much smaller than the micro-stick-slip period $T_{\rm mss}$, in agreement with our measurements. 
 Therefore, in this approach,
 the fracture kink transverse propagation proceeds 
 at the  group velocity  $v_g = 2 \lambda /\tau$ of bending waves of wavelength $\lambda = A_{\rm mss}$.
  The estimated velocities  ($650\rightarrow 900$~m/s) are also in excellent agreement with the experimental reported values~\cite{Dalbe2015}.

On the other hand,  the increase of kinetic
energy  that the tape locally gains during 
a micro-slip is  $E_k = \frac12 \mu A_{\text{mss}}
\left(A_{\text{mss}}/T_{\text{mss}}\right)^2 2(1-\cos\theta)$.
In this formula, the factor
$2(1-\cos\theta)$ results from the motion of the tape just beyond the curved region, which is a combination of  
a translation 
in the direction $\theta$ at $V_{\text{mss}}= A_{\text{mss}}/T_{\text{mss}}$ and a translation at the
same velocity in the direction of the peel front motion
\cite{Dalbe2016}. 
Finally, the transfer of bending to kinetic energy $E_{B-r}  = E_k $ allows
to link amplitude and period of the micro-stick-slips: 
\begin{equation}\label{eq:AT}
A_{\text{mss}} = \left[ \frac{2 B/\mu}{1-\cos\theta}\right]^{1/6}
d^{1/3} \ T_{\text{mss}}^{1/3}.
\end{equation}
The predictions of Eq.~(\ref{eq:AT}) are in excellent
agreement with our measurements: 
the power-law exponent of $1/3$ between
$A_{\text{mss}}$ and $T_{\text{mss}}$, the independence with the
peeled length $L$ 
and the weak dependence with the peel angle $\theta$ and ribbon bending modulus $B$ through the $1/6$ exponent  
reproduce 
the various measured dependencies reported in 
Figs.~\ref{fig:L_Theta_B} and~\ref{fig:BK}. 
Indeed,
normalizing 
the amplitude $A_{\text{mss}}$ 
by the 
pre-factor  $\left[ 2B/ (\mu (1-\cos\theta))\right]^{1/6}$, 
in Eq.~(\ref{eq:AT}), which accounts for the changes in mass
and bending modulus of the ribbon, as well as the different peel angle used,  
tends to gather the data on a master curve.   
The data collapse is  particularly convincing for the samples with different backing thickness as shown in
Fig.~\ref{fig:BK}.  
Moreover, 
following Eq.~(\ref{eq:AT}), we could fit the large amount of data reported in Fig.~\ref{fig:L_Theta_B}, 
for the various experiments performed at different $(L,\theta)$ with only one backing, 
to extract the free parameter $d$, corresponding to the critical elongation at which the glue detaches, $d = 5$~$\mu$m.  
Such order of magnitude is compatible with our direct side observations of the adhesive 
tape unstable peeling.
Strikingly, with this single value for $d$, we can finally  describe quantitatively our various experiments 
with samples of different lineic mass $\mu$ and bending modulus $B$, as shown by the dashed line reported 
in  Fig.~\ref{fig:BK}.

To conclude, thanks to an extensive experimental study in addition
to a careful preparation of adhesive-substrate joints with the
exploration of several tape backing bending modulus, we have been
able to unveil the precise characteristics of the detachment front
micro-stick-slip dynamics, appearing when peeling an adhesive tape
at high velocities. 
A local energy balance of the detachment process 
shows that  
the elastic bending energy stored in the tape region
that will detach during the micro-slip is converted into a kinetic energy
increase of the peeled tape during a micro-stick-slip cycle.
Our model  allows a quantitative description of 
 the observed scaling-law linking  
amplitudes and periods of the micro-instability, and in particular its dependency with the
peeling angle, as well as with the bending modulus and lineic mass of the ribbon.  
This energy transfer arising from the assumption of the complete dissipation of the energy stored in the deformation of the adhesive layer when it detaches, by elastic hysteresis,    
highlights that the rapid micro-slip corresponds to a specific dynamic rupture mode of propagation of the detachment front. 
In our scenario, 
the elastic stretching energy stored in the whole peeled ribbon during the stick phase of the macro-instability and released during the macro-slip phase constitutes  an energy reservoir for the micro-stick-slip cycles and more precisely for the reloading of the tape local bending during the micro-sticks. The release of the stretching energy therefore proceeds by quanta of elastic bending energy of the ribbon close to the peel front.

Nevertheless, the physical origin of the kinked detachment front propagation 
in the direction transverse to the main peeling direction
still needs to be uncovered. A possible explanation could come from a local enhancement of the mechanical energy release rate, which has been shown~\cite{AddaBedia2006, Vilmin2010} to be at the origin of the elastic fingering instability when peeling quasi-statically a confined elastomer~\cite{Ghatak2000}.  

\acknowledgments  The  ANR
Grant ``AdhesiPS'' No.~ANR-17-CE08-0008, the MegaGrant No.~14.W03.31.0002 and the LIA ``D-FFRACT" supported this work.

\end{document}